\documentclass[prl,showpacs,preprintnumbers,twocolumn,amsmath,amssymb,superscriptaddress]{revtex4-1}

\usepackage[T1]{fontenc}
\usepackage[applemac]{inputenc}
\usepackage[english]{babel}
\usepackage{graphicx}
\usepackage{amssymb}
\usepackage{amsmath}
\usepackage{overpic}
\usepackage{color}

\newcommand{\be}{\begin{equation}}
\newcommand{\ee}{\end{equation}}
\newcommand{\bea}{\begin{eqnarray}}
\newcommand{\eea}{\end{eqnarray}}

\def\m{\mu}

\def\G{\Gamma}
\def\D{\Delta}

\def\bQ{{\bf Q}}

\def\lb{\label}

\newcount\bozza \bozza=0
\ifnum\bozza=1
\newdimen\shift \shift=-2truecm
\def\lb#1{%
{\label{#1}\rlap{\kern\shift{$\scriptstyle#1$}}}}
\else\def\lb#1{\label{#1}} \fi

\newcommand{\Nb}{{$2H$-NbSe$_2$ }}

\begin{document}

\title{Lattice contribution to the unconventional charge density wave transition in $2H$-NbSe$_2$: a non-equilibrium optical approach}
%Unconventional charge density wave physics and Lattice Degrees of Freedom in \Nb: a Non-equilibrium Optical Approach
%Disentangling electronic and lattice contributions to the unconventional charge-density-wave transition of \Nb by non-equilibrium optical spectroscopy
%Electronic and lattice contributions to the unconventional charge-density-wave transition of \Nb revealed by non-equilibrium optical spectroscopy
%Robustness of the periodic lattice distortion in the charge density wave material \Nb
%Disentangling the electronic and lattice degrees of freedom in the charge-density-wave phase of 2H-NbSe2 by non-equilibrium optical spectroscopy.
%Disentangling the electronic and lattice degrees of freedom in the charge-density-wave material 2H-NbSe2 by non-equilibrium optical spectroscopy.
%Role of the electronic and lattice degrees of freedom in the unconventional charge-density-wave transition in 2H-NbSe2

\author{Daniel T. Payne}
\affiliation{Elettra - Sincrotrone Trieste S.C.p.A., 34149 Basovizza, Trieste, Italy}
\author{Paolo Barone}
\affiliation{Istituto Superconduttori, Materiali Innovativi e Dispositivi (SPIN-CNR), c/o Universit{\`a} G. D'Annunzio, I-66100 Chieti, Italy}
\author{Lara Benfatto}
\email{lara.benfatto@roma1.infn.it}
\affiliation{ISC-CNR and Dep. of Physics, Sapienza University of Rome, P.le A. Moro 5, 00185 Rome, Italy}
\author{Fulvio Parmigiani}
\affiliation{Elettra - Sincrotrone Trieste S.C.p.A., 34149 Basovizza, Trieste, Italy}
\affiliation{Department of Physics, Universit\`a degli Studi di Trieste, I-34127 Trieste, Italy}
\affiliation{International Faculty, University of Cologne, Albertus-Magnus-Platz, 50923 Cologne, Germany}
\author{Federico Cilento}
\email{federico.cilento@elettra.eu}
\affiliation{Elettra - Sincrotrone Trieste S.C.p.A., 34149 Basovizza, Trieste, Italy}

\date{\today}

\begin{abstract}
The complex Fermi surfaces of transition-metal dichalcogenides (TMDCs) challenge the standard Peierls-instability-driven charge-density-wave (CDW) formation. Recently, evidence has been accumulating of a prominent role of ionic thermal fluctuations, which frozen out below $T_{CDW}$ inducing a periodic lattice distortion (PLD). We focus on $2H$-NbSe$_2$, displaying a quasi-commensurate CDW below $T_{CDW}$ $\approx$33 K, and use time-resolved optical spectroscopy (TR-OS) to detect and disentangle the electronic and lattice degrees of freedom. We reveal a fingerprint of the ordered phase at h$\nu$ $\sim$1 eV: at $T_{CDW}$, a divergent relaxation timescale and a sign-change of the differential reflectivity indicate that CDW gap opening and PLD formation occur at the same temperature. However, we show that these effects can be decoupled under moderate photoexcitation, forming a long-lived state in which the electronic order is destroyed, but the lattice distortion is not. Our results and computations suggest an unconventional CDW mechanims in $2H$-NbSe$_2$, highlighting the dominant role of the lattice in driving the ordered-phase formation.
\end{abstract}
%\pacs{74.20.-z,74.25.Gz,74.25.N-}

\maketitle

{\em INTRODUCTION}

Transition-metal dichalcogenides (TMDCs), such as NbSe$_2$, TaSe$_2$ and TaS$_2$, warrant considerable scientific attention due to their exotic physical properties. A common behavior is the appearance, below a material-dependent temperature, of a charge-density-wave (CDW) modulation, sometimes followed by a superconducting order \cite{rossnagel_review_2011}. 
The simplest Peierls mechanism attributes the CDW formation to an instability of the Fermi surface, driven by the existence of electron-like and hole-like states connected by the so-called nesting wavevector ${\bf Q}$. In this situation, the electronic susceptibility at ${\bf Q}$ is strongly enhanced, and the coupling to phonon modes at the same nesting wavevector can trigger the spontaneous breaking of symmetry below a critical temperature ($T_{CDW}$), inducing simultaneously  a spatially-periodic lattice distortion (PLD) and a modulation of the charge density (CDM). In one dimension the nesting condition is always fulfilled for $\bQ=2k_F$, where  $k_F$ is the Fermi wavevector,  making a Peierls-like CDW instability possible even for weak electron-phonon coupling \cite{gruner_review}. A similar phenomenon can occur in higher dimensional systems if the nesting condition still occurs, as e.g. when large regions of the Fermi surface lie parallel to one another. In this case the CDW transition is expected to open a single-particle gap at the Fermi surface at the regions connected by the nesting wavevector.

Even though some indications of a nesting-induced gap opening has been evidenced by ARPES \cite{arpes1,arpes2,arpes3,arpes4,arpes5,straub_1999}, the mechanism underlying the CDM/PLD formation in many TMDCs appears more complex, with remarkable differences between the various families \cite{rossnagel_prb06,cavalleri_prl11,bauer_nature11}. RETe$_3$ systems, in which RE is a rare earth ion, are ideal models of weak electron-boson coupled CDW formation, where the gap at $E_F$ is induced by nesting \cite{yusupov_prl08}. In pump-probe measurements of TbTe$_3$, the gap at $E_F$ closes on the timescale of structural distortions, implicating electron-phonon coupling as the underlying origin \cite{shen_science08}. However, the Fermi surfaces of some TMDCs are not well-nested in geometry \cite{rossnagel_review_2011}. Unlike RETe$_3$, electron-phonon coupling in these materials is strong \cite{rossnagel_review_2011} and the nesting condition is barely fulfilled, challenging the assumption that CDW formation is driven solely by a canonical Peierls mechanism. Approaching the problem from an out-of-equilibrium perspective can allow one to discern the nature of the dominant type of interaction behind CDW formation, as well as the order parameters, as described in \cite{hellmann_2012}. Our aim is to unveil these mechanisms in $2H$-NbSe$_2$.

In this work we use time-resolved optical spectroscopy (TR-OS) to address the nature of the CDM/PLD formation in bulk 2H-NbSe$_2$. By measuring the reflectivity variation as a function of the pump-probe delay we identify clear signatures of the CDW transition at energies far away from $E_F$. Interestingly, despite the fact that the gap opens in selected regions of the FS, our pump pulse at 1.55 eV strongly affects the electronic states. This can be explained by ab-initio calculations showing that the optical transitions at this energy are strongly momentum selective, and mainly affect the electronic states with a large CDW gap. We track the occurrence of the phase transition at quasi-thermal equilibrium across $T_{CDW}$; in this situation long-range order is lost by increasing the temperature and the high-temperature phase is reached. Conversely, by increasing the pump fluence we can trigger and disentangle the destruction of CDM and PLD orders separately. Indeed, we detect the signature of the melting of the CDM electronic order at a critical fluence, while the PLD response persists to higher fluences. Hence, we photo-induce a decoupling of the lattice and electronic contributions to the CDM/PLD phase, such that the PLD persists while electronic charge ordering is reduced. The additional energy required to destroy the PLD suggests a phononic-driven origin of the CDW transition, providing experimental support to an unconventional mechanism of CDW formation in 2H-NbSe$_2$.

In bulk $2H$-NbSe$_2$ the transition to a near-commensurate CDW phase occurs at $T_{CDW}$ $\approx$33 K \cite{harper_1975, moncton_1975}. The mechanism driving CDW formation has been matter of intense debate \cite{mazin_prb06,mauri_prb09,zhu_pnas15,flicker_natcomm15,rodiere_prb15,Weber_prb16,flicker_prb16,mauri_cm20}, further triggered by recent measurements in monolayers \cite{mak_natnano15,staley_2009,Ugeda_NP_2015,lin_cm20}. In monolayer samples, the $T_{CDW}$ is enhanced up to 145 K, while the superconducting phase, which coexists with the CDW below $\approx$7 K in the bulk, is suppressed. The magnitude of the CDW-induced gap at $E_F$ is also controversial. The Fermi surface (FS) of $2H$-NbSe$_2$ consists of double-walled barrels around the center ($\G$ point) and corners (K) points of the Brilluoin zone, as well as a small three-dimensional pancake FS around the $\G$ point. Direct ARPES measurements suggest that the gap at $E_F$ opens mainly in the quasi-2D Fermi sheets around the K point, which exhibit Nb 4d-like character, with an average CDW gap value ranging from 2 to 5 meV, the maximum being located along the K-M direction \cite{arpes1,arpes2,arpes3,arpes4,arpes5}. This would be roughly consistent with the hot-spot locations of the CDW nesting wavevectors, where $Q\approx (2\pi/3,0)$. However, the opening of the gap in isolated regions of the FS challenges the usual FS-driven Peierls mechanisms, since such a small gain in electronic energy associated with the CDM cannot overcome the cost of introducing the PLD associated with CDW order. This view is further supported by tunnelling experiments, which reveal a significant spectroscopic variation in the density of states only $\sim$0.7 eV below $E_F$ \cite{cava_prb14}. All these findings suggest that the CDW in NbSe$_2$ is mainly triggered by phononic degrees of freedom, and that anharmonic phonon-phonon interactions play a major role, as evidenced by ab-initio and field-theory calculations \cite{mauri_prb09,flicker_natcomm15,flicker_prb16,mauri_cm20}. In addition, by taking into account the strong momentum dependence of the electron-phonon coupling one can recover the enhancement of the electronic susceptibility at the nesting wavevectors \cite{mauri_prb09,flicker_natcomm15,flicker_prb16}, explaining the observed momentum dependence of the CDW gap.

\begin{figure}[htb]
\centering
\includegraphics[width=0.48\textwidth,angle=0]{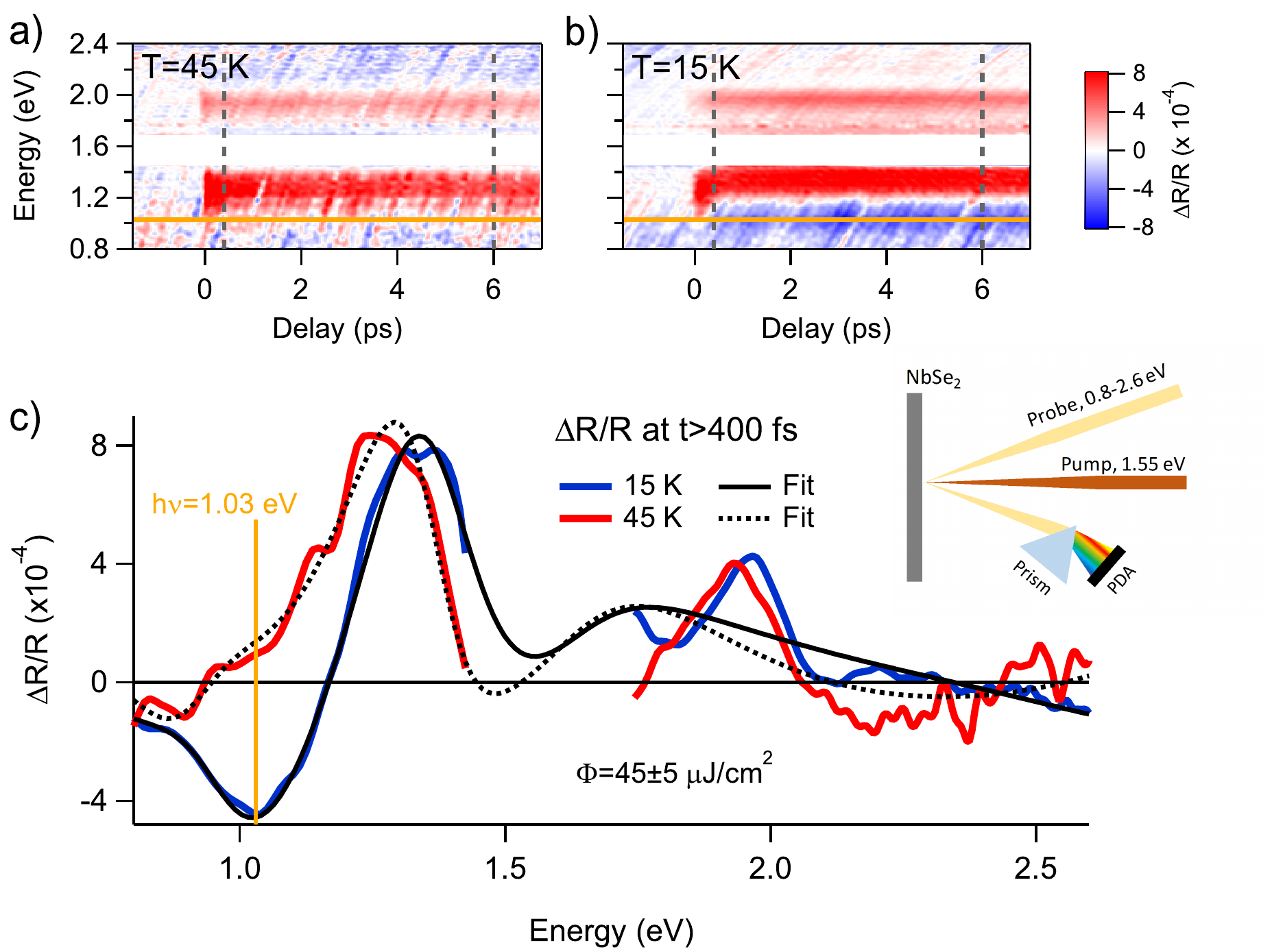}
\caption{Time-and-frequency resolved normalized reflectivity variations, $\D R/R(t,h\nu)$, of 2H-NbSe$_2$, measured at (a) 45 K and (b) 15 K. The colorscale is indicated on the right. Measurements have been acquired at temperatures above and below the CDW transition temperature T$_{CDW}$=33 K, respectively. The region around the pump photon energy (1.55 eV) has been blanked because the signal was compromised by diffusion of pump photons from the sample surface. The experimental setup used is also sketched. (c) Spectral profiles at 15 K (in blue) and 45 K (in red), integrated between the two vertical dashed lines (400 fs < $t$ < 6 ps) drawn in panels (a) and (b). Black lines (solid and dashed) are the fit to the data, performed as described in the main text. The orange line in the panels denote the photon energy (1.03 eV, or 1200 nm) at which single-color probe measurements have been performed. At this photon energy, the CDW phase displays its most marked fingerprint in the $\D R/R(h\nu)$ signal, with a sign-change in the $\D R/R(h\nu)$ across the CDW-to-normal-phase phase transition, in the near-infrared spectral range.}
\label{fig1}
\end{figure}

\vspace{0.5cm}
{\em RESULTS}

Time-and-frequency resolved reflectivity measurements (see Materials and Methods) were conducted for sample temperatures above and below $T_{CDW}$, revealing the out-of-equilibrium response of the material over a broad spectral range (0.8-2.6 eV). Measurements were performed at a fluence of 45$\pm$5 $\mu$J/cm$^2$, pumping at 1.55 eV. The $\D R/R(t,h\nu)$ signal remains essentially unchanged as the sample is cooled from room temperature to slightly above $T_{CDW}$. Figure 1a) shows a $\D R/R(t,h\nu)$ map acquired at T=45 K. However, below $T_{CDW}$, a large and negative component in the $\D R/R(h\nu)$, centered at $h\nu\sim$1 eV and $\sim$0.2 eV wide sets in after approximately 400 fs. This finding is shown in Figure 1b), displaying a $\D R/R(t,h\nu)$ map acquired at T=15 K. The negative signal is not observed at any temperature above $T_{CDW}$. Following the instantaneous response, a quasi-equilibrium is reached after $\sim$400 fs and persists for tens of picoseconds. Similar transient behaviour of the CDW/PLD has been observed in 2H-TaSe$_2$ \cite{wang_apl13}. The quasi-equilibrium state evolves on a timescale characteristic of lattice thermalization.

The transient responses of $2H$-NbSe$_2$ in the CDM/PLD and high-temperature phases are compared in panel (c), which displays the signals shown in Figure 1(a) and (b) integrated in the interval 0.4-6 ps. The blue curve represents the $\D R/R(0.4-6 $ ps$)$ at 15 K, the red curve the $\D R/R(0.4-6 $ ps$)$ at 45 K, while the black lines are the fit to the curves, obtained as described in the following. Due to the requirement to operate at low fluence, the $\D R/R$ signal is in the 10$^{-4}$ range,  hence at the signal-to-noise limit of our InGaAs detectors. Nonetheless, we unambiguously demonstrate that the onset of the CDW phase in 2H-NbSe$_2$ has a clear fingerprint in the near-infrared spectral region at $\approx$1 eV. This range is not commonly investigated in time-resolved experiments, thus explaining why the evidence of the CDW phase in out-of-equilibrium experiments on 2H-NbSe$_2$ remained elusive so far \cite{anikin_2020}. However, this behavior is consistent with the evolution of the optical properties at equilibrium as a function of temperature (\cite{basov_prb01} and Supplementary Materials), where the largest variation is indeed peaked at $\approx$1 eV. Remarkably, in the out-of-equilibrium experiments, the sign of $\D R/R(t>$400 fs$,\approx1$ eV$)$ changes across $T_{CDW}$; this fact will be exploited to track and study the behavior of the temperature-induced and fluence-induced phase transitions, by high-resolution single-colour reflectivity experiments, as we will explain in the rest of the paper.

In order to analyze the $\D R/R(h\nu)$ transients acquired at 15 K and 45 K, we proceed as follows. A model of the in-plane equilibrium reflectivity at 10 K was created on the basis of data presented in Reference \cite{basov_prb01}. It consists of four Lorentz oscillators (centered at 1.37 eV, 1.63 eV, 1.96 eV, 3.23 eV) and two Drude terms (a broad one and a narrow one), having similar weight. The details of this analysis are reported in the Supplementary Materials. The out-of-equilibrium signal was emulated by slightly modifying a few parameters of the model dielectric function, and calculating the simulated $\Delta R/R(h\nu)$ response. Both at 15 K and 45 K, a reasonable agreement between data and model is obtained by allowing the Lorentz oscillators at 1.37 eV (11040 cm$^{-1}$) and at 1.96 eV (15790 cm$^{-1}$) to vary, as well as the broad Drude component. At both temperatures, we observe a reduction of the plasma frequency $\omega_p$ of the Lorentz oscillator at 1.96 eV, and an increase of the plasma frequency of the Lorentz oscillator at 1.37 eV. Overall, the spectral weight of the high-energy, interband transitions, SW$_{inter}$=(1/8)$\sum_{i=1}^{4} \omega^2_{p,i}$, is reduced when the system is brought out-of-equilibrium, that is, $\delta$SW$_{inter}\equiv$ SW$_{inter,non-eq}$-SW$_{inter,eq}$<0. However, in the CDM/PLD phase, $\delta$SW$_{inter}$ is twice as large than the value obtained in the high-temperature phase. This finding indicates that the quench of the CDM/PLD phase by the pump pulse implies a marked modification of the spectral weight at energy scales far larger than the energy of the electronic gap. This spectral weight is transferred to spectral regions outside the photon energy range we probe. Such a large redistribution of spectral weight is reminiscent of what observed in cuprate superconductors \cite{giannetti_2011,giannetti_review}, and it is a first indication of an unconventional CDW mechanism. Indeed, the fact that the CDM/PLD phase affects the optical properties at energies as high as $\sim$1 eV, a value well in excess than the CDW gap in 2H-NbSe$_2$, reveals the presence of a complex interplay of states at different energy scales, that require further investigation. As an example, in the case of cuprate superconductors, the link between states at much different energy scales is set by the Mott-like physics \cite{giannetti_2011}.

\begin{figure*}[htb]
\centering
\includegraphics[width=0.9\textwidth,angle=0]{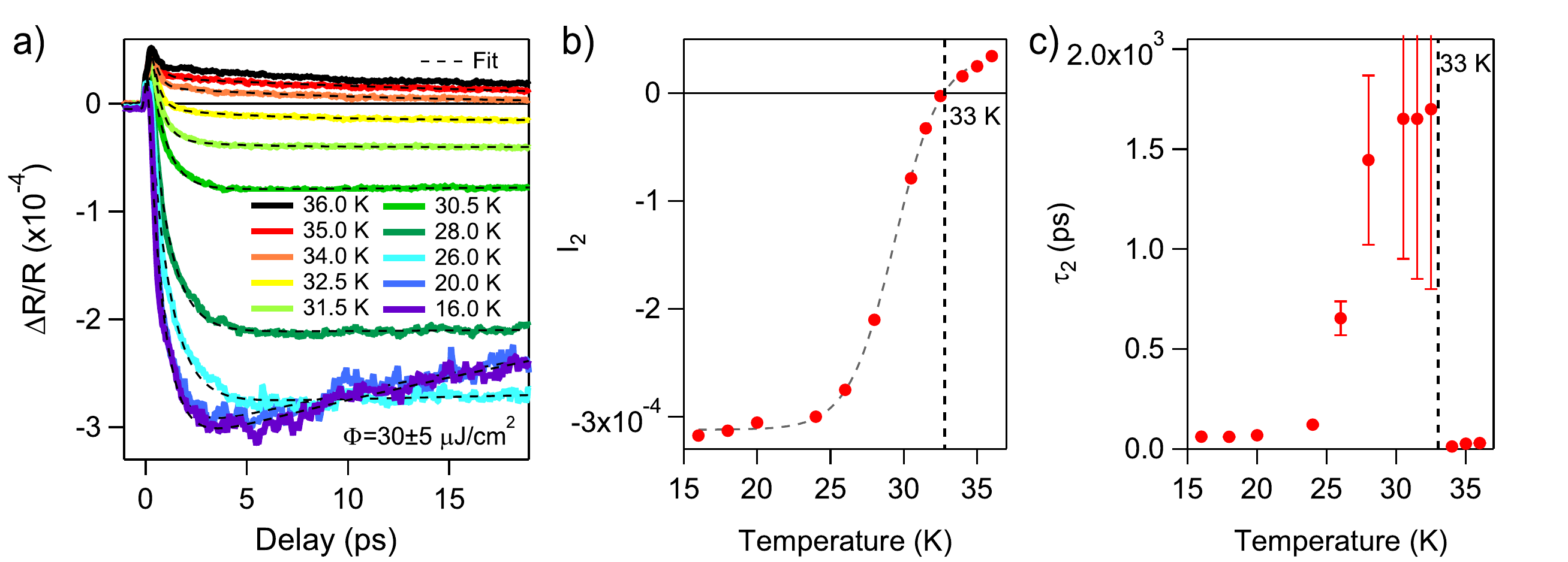}
\caption{Temperature dependence of the single-colour probe spectra at $h\nu$=1.03 eV and 30$\pm$5 $\mu$J/cm$^2$ fluence. (a) The $\Delta R/R(t)$ signal is satisfactorily fitted with the convolution of two exponential decays, representing an instantaneous and a slowly-evolving contribution, respectively. (b) Amplitude of the slow-component ($I_2$) as a function of temperature. The dashed line is presented as a guide to the eye. (c) Decay time of the slow-component ($\tau_2$). The transition temperature $T_{CDW}$=33 K is indicated in panels (b) and (c). The large error bars accompanying the values of $\tau_2$ in the vicinity of the transition temperature from below are explained by the fact that the values of $\tau_2$ are large compared to the time-window investigated.}
\label{fig3}
\end{figure*}

In order to better understand the evolution of the CDM/PLD-induced modifications, we performed single-colour probe measurements at the photon energy where $\Delta R/R(h\nu)$ varies more markedly across $T_{CDW}$. Hence, we tuned the probe at 1.03 eV (equivalent to 1200 nm or $\sim$8300 cm$^{-1}$), as marked by the orange line in Figure 1. These curves, reported in Figure 2(a), show positive $\D R/R$ above $T_{CDW}$ and negative $\D R/R$ in the CDM/PLD phase after $\sim$400 fs. Hence, we identify the negative, quasi-equilibrium signal at 1.03 eV as a diagnostic fingerprint of charge-ordering in 2H-NbSe$_2$. This transient response is typical of CDW systems \cite{demsar_prb14}. We stress the fact that the $\D R/R$ signal in the vicinity of the transition temperature displays extremely large decay timescales, indicating the realization of a quasi-equilibrium condition for the excited quasiparticles due to the blocking of the available decay channels. 
The single-colour spectra were fit by a convolution of two exponential decays, representing an instantaneous and a slowly-evolving decay channel, with a Gaussian representing the experimental time resolution due to the pump-probe cross-correlation (having Full Width at Half Maximum, FWHM, of 100 fs). We denote with $I_1$, $\tau_1$ and $I_2$, $\tau_2$ the amplitude and time-constant of the fast and slow exponential components, respectively.  
The values of $I_2$ and of $\tau_2$ as obtained from the fitting procedure are shown in Figure 2(b) and (c), respectively.
The amplitude of the quasi-equilibrium signal $I_2$ changes sign at $\sim$33 K, in agreement with the known $T_{CDW}$ \cite{harper_1975, moncton_1975}. Similar transient behaviour has previously been observed in several systems, including the normal-to-superconducting phase transition in over-doped Y-Bi2212 \cite{parmigiani_prb11}. A concomitant divergence in the relaxation time (quasiparticle-lifetime) when the transition is approached from below is identified, indicative of a second-order phase transition \cite{demsar_prb02, giannetti_review}. This effect has been observed in charge-density wave systems as well as in several superconducting compounds, including MgB$_2$, copper-based superconductors and iron-based superconductors \cite{demsar_prb02, Kuo_2019, Demsar_2003, Han_1990, Kabanov_1999, Demsar_1999, Dvorsek_2002, Kabanov_2005, Chia_2010}. This phenomenon is often interpreted as an indication that the excitation gap closes and a collective order in the system is destroyed. 
For the case of cuprate superconductors, it was possible to quantitatively extract the gap amplitude from the characteristic relaxation time divergence at the transition \cite{Kabanov_1999}. In general, the divergence of the relaxation time at the critical temperature $T_c$ is indicative of the progressive decrease of the phase-space available to inelastic scattering processes involving a fraction $\Delta(T)/(k_BT_c)$ of the thermally occupied states. The precise value of the relaxation time depends on the microscopic details, but typically is proportional to the inelastic scattering time of a quasiparticle at the Fermi surface.
However, as discussed in the introduction, in 2H-NbSe$_2$ below $T_{CDW}$ only a small fraction of the FS is modified by the CDW distortion, with the gap opening only in selected k-space regions, and indeed the system remains metallic, challenging the understanding of the strong influence of the CDW gap on the $\Delta R/R$ signal. 

\begin{figure*}
\centering
\includegraphics[width=0.9\textwidth]{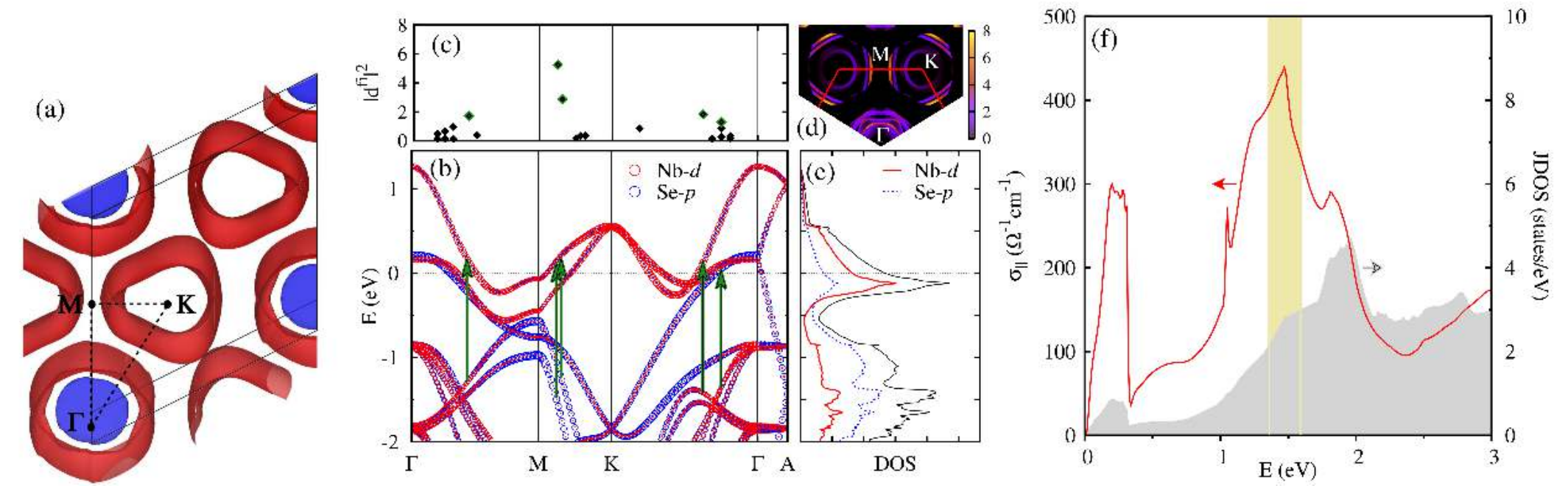}
\caption{(a) Fermi surface of \Nb in the normal state. (b) Calculated band structure of \Nb with the atomic contributions of Nb-$d$ (red) and Se-$p$ (blue) states proportional to the size of the coloured dots.  Arrows denote the location of the interband transitions with the largest dipole matrix elements $\vert d^{fi}\vert^2$ between an initial (valence) state $i$ and a final (conduction) state $f$ with an energy transfer $\Delta E=\varepsilon_{f\bm k}-\varepsilon_{i\bm k}$ ranging from 1.38 $eV$ to 1.58 $eV$, corresponding to the energy interval where the interband optical conductivity displays a strong peak (see panel (f)). The calculated $\vert d^{fi}\vert^2$ within the chosen energy window are shown in arbitrary units along the high-symmetry line $\Gamma$-M-K-$\Gamma$-A (c) and in the reciprocal-space plane $k_z=0$ (d), highlighting its pronounced momentum dependence. (e) Density of states alongside its decomposition in atomic-orbitals contribution, using the same color code as in (b). (f) Calculated interband contribution to the (in-plane) optical conductivity $\sigma_\parallel$ of \Nb in the normal state, alongside with the JDOS, shown as a shaded area. The coloured area denotes the energy window centered at the position ${E}_{max}=1.48~eV$ of the maximum peak in $\sigma_\parallel$, selected for the calculations of the dipole matrix elements shown in panels (d),(e).}
\label{fig-theo}
\end{figure*}

To elucidate this issue we performed ab-initio calculations (see Materials and Methods) of the optical conductivity of 2H-NbSe$_2$ resolved in momentum space, in order to identify the relevant electronic transitions triggered by the pump pulse. As shown by measurements of the equilibrium reflectivity on 2H-NbSe$_2$ \cite{basov_prb01}, the optical properties at the pump frequency do not change significantly below $T_{CDW}$, so that we can safely infer the effect of the pump by ab-initio calculations in the non-CDW phase. In Fig.\ \ref{fig-theo} we show the band structure of 2H-NbSe$_2$ along with the main (Nb or Se) character of the density of states (DOS).  As one can see, even though states at $E_F$ have mainly Nb character, the mixing with Se-originated bands is crucial in order to guarantee a finite probability for interband transition. In addition, we find a pronounced momentum dependence of the dipole moments, which are larger at the wavevector marked by arrows in the figure. In particular, for an energy difference of 1.48$\pm$0.1 eV, one observes a strong enhancement along the $K-M$ direction, where the CDW gap opens below $T_{CDW}$ \cite{rahn_gaps_2012}. This effect results in the strong absorption peak centered around 1.48 eV in the interband optical conductivity, shown in Fig.\ \ref{fig-theo}f. As one can see, direct comparison with the joint DOS, which does not include the effect of the optical matrix elements, further confirms the momentum-selective nature of the process. We can then infer that a pump pulse at 1.55 eV affects predominantly the electronic states where the largest CDW gap opens below $T_{CDW}$, explaining why the out-of-equilibrium dynamics reported in Figure 2 closely resembles the one observed in fully-gapped CDW or superconducting systems.

\begin{figure*}[htb]
\centering
\includegraphics[width=\textwidth,angle=0]{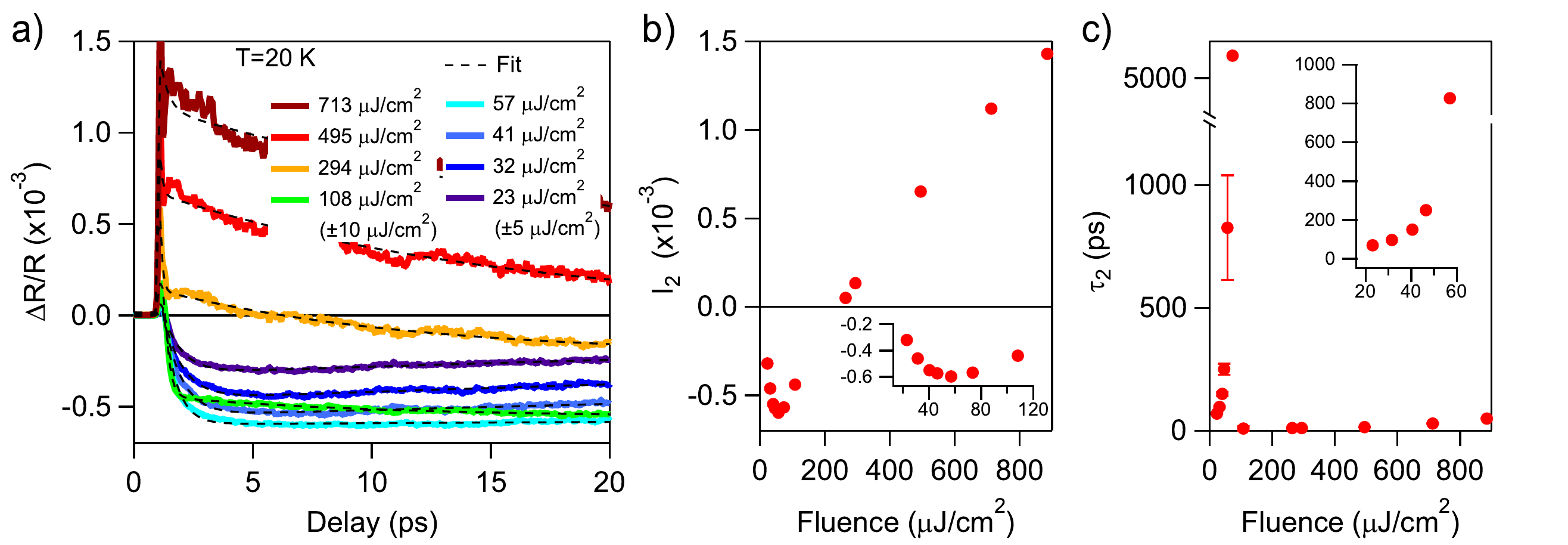}
\caption{Fluence dependence of the single-colour probe spectra (h$\nu$=1.03 eV) measured at T = 20 K. (a) Spectra as a function of the pump-probe delay. Black dashed lines are the fit to the curves, performed with a double exponential decay. (b) Amplitude of the quasi-equilibrium fitting parameter, $I_2$. (c) Decay Time of the quasi-equilibrium fitting parameter, $\tau_2$. The insets show more clearly the maximum of the amplitude and the divergence of the decay time at a fluence of $\sim$60 $\m$J/cm$^2$.}
\label{fig4}
\end{figure*}

Once identified the main signatures of the CDW formation at thermal equilibrium we further explore the ability of ultra-fast photoexcitation to disentangle electronic (CDM) and phononic (PLD) contributions. In Fig.\ \ref{fig4} we report the dependence of the $\D R/R$ signal at 1.03 eV on the absorbed pump fluence for $T=20$ K. In the weakly-perturbative regime, the signal amplitude is negative at late times and increases in magnitude until $\sim$60 $\m$J/cm$^2$ (see Fig.\ \ref{fig4}b and its inset), where the decay time diverges (see Fig.\ \ref{fig4}c and its inset). 
In the thermally-induced phase transition, CDM and PLD are lost simultaneously at $T_{CDW}$. Interestingly, here we observe still a negative $I_2$ after the vaporization of the long-range electronic order, indicated by the divergence of $\tau_2$ at $\sim$60 $\m$J/cm$^2$. Hence, we can affirm that it is possible to disentangle the electronic and lattice distortions, the latter being identified by a negative $I_2$ (at h$\nu$=1.03 eV). The decoupling of these features between $\sim$60 $\m$J/cm$^2$ and $\sim$250 $\m$J/cm$^2$ suggests that a long-lived, non-equilibrium photoinduced phase is accessible in which long-range electronic order is destroyed, yet the PLD remains. Only at fluences larger than $\sim$250 $\m$J/cm$^2$, the $I_2$ component turns to positive and also the PLD is destroyed.

\vspace{0.5cm}
{\em DISCUSSION}

Our finding is at odd with respect to recent results obtained on underodped cuprate superconductors, where the photoinduced vaporization of  superconductivity is not accompanied by a lifetime divergence at the critical fluence \cite{giannetti_prb09, parmigiani_prb11}. This apparent contradiction can be rationalized by noting that in underdoped cuprates the small energy scale driving the transition is set by the phase stiffness. As a consequence the ultrafast collapse of superconductivity happens by quenching of the phase coherence rather than the pairing of Cooper pairs \cite{boschini_2018}, hence no lifetime-divergence can be observed. Conversely, in the case of NbSe$_2$ our results suggest that the charge-density-wave gap, which opens in the electronic spectrum, is the lower energy scale. As a consequence the photoexcitation directly quenches the electronic CDW order without affecting the gain in energy of the lattice structure via the PLD. The comparison between the thermal-induced and the fluence-induced transition supports this picture. While at quasi-equilibrium the divergence of the quasi-particle lifetime occurs concomitantly with the change in sign of the $\D R/R$ amplitude, in the photoinduced transition the two phenomena can be decoupled. The destruction of the CDW gap, signalled by the divergence of the quasiparticle relaxation time, occurs already at $\sim$60 $\m$J/cm$^2$, where a large negative $\D R/R$ signal is still present, indicative of a well formed PLD. At this fluence, the pump-induced heating is estimated to be <7 K, and the sample remains comfortably below $T_{CDW}$. Additionally, since the broadband transient response of the system in this fluence regime could not be modelled by a linear combination of the signals measured above and below $T_{CDW}$ (Figure 1), we exclude the formation of metallic domains as a possible mechanism.
Above this critical fluence the density of photoexcited carriers is sufficiently large to perturb also the PLD order and subsequently, the signal becomes less intense. Finally, the quasi-equilibrium $\Delta R/R$ signal component becomes positive above $\sim$250 $\m$J/cm$^2$ and a linear dependence is recovered. At this fluence, it is hard to establish whether the quench of the PLD is simply due to the average heating due to the pump pulse power, or if this is a non-equilibrium effect too. Interestingly, the intensity of the spectrum measured at 294 $\m$J/cm$^2$ changes sign after $\sim$5 ps. This provides a possible insight into the timescales required for the recovery of the PLD.
The photo-induced decoupling of the electronic CDM and the PLD, which are strongly coupled at equilibrium, has also been observed in 1T-TaS$_2$ \cite{rossnagel_prl10}. The persistence of the PLD following the non-thermal destruction of long-range CDW order, suggests that the PLD may constitute the dominant contribution to the energy gain associated with low-temperature instability in 2H-NbSe$_2$. This observation is well supported by ab-initio and field-theory calculations \cite{mauri_prb09,flicker_natcomm15,flicker_prb16,mauri_cm20} pointing to a relevant role of anharmonic phonon interactions, consistently with recent reports on the CDW evolution under pressure in 2H-NbSe$_2$ \cite{rodiere_prb15}.

Finally, we notice that coherent oscillations associated with the CDW amplitude mode are commonly observed in time-resolved studies of similar systems \cite{rettig_natcomm16,shen_science08, demsar_prb14,demsar_prb02,wolf_prb16,perfetti_njp08}, which are not revealed in this study. However, as shown by direct equilibrium Raman measurements \cite{tsang_prl76,measson_prb14,measson_prb18}, which are well reproduced theoretically by taking into account the momentum dependence of the gap \cite{cea_cdw_prb14}, the amplitude mode in 2H-NbSe$_2$ is rather broader than in other systems, and it only develops a clear resonance at temperatures significantly lower than $T_{CDW}$. At the lowest temperature accessible in our measurements the soft-phonon mode is expected to lead to strongly overdamped oscillations \cite{Udina_2019}, which can be masked by the electronic depairing processes induced by interband transitions.

In summary, we have performed time-resolved optical spectroscopy measurements on the CDW system 2H-NbSe$_2$ and we revealed a clear fingerprint of the onset of the CDW phase in the $\D R/R$ signal at high-photon-energy. We identified clear electronic and pohononic signatures of the CDM/PLD phase transition. Despite the fact that in 2H-NbSe$_2$ the gap opens in restricted regions of the FS, making it difficult its identification via conventional spectroscopies at equilibrum, our pump-probe approach meets the conditions for its detection. Indeed, as we showed by direct ab-initio calculations of the absorption spectrum, a pump pulse at $\sim$1.55 eV selectively triggers interband transitions at the momenta where the CDW gap is larger. At equilibrium, the onset of long-range electronic CDM order and the PLD are inextricably linked, and they occur at the same temperature, $T_{CDW}$. However, we have shown that CDM and PLD may be decoupled non-thermally by photoexcitation, showing the robustness of PLD against the photoinduced melting of the electronic charge ordering, in strong contrast to the canonical Peierls mechanism where the electronic instability drives the transition. Growing evidence of the fact that a PLD can be responsible for the CDW transition by providing the main source of energy stabilization is emerging, as recently demonstrated in particular for single-layer compounds in \cite{silva_2016}. Our results strengthen the emerging picture of an unconventional CDW transition in TMDCs, where anharmonic phonon-phonon interactions and strong momentum-dependent electron-phonon coupling give rise to a novelty of correlated electronic states.

\vspace{0.5cm}
{\em MATERIALS AND METHODS}

\textbf{Time-Resolved Optical Spectroscopy}

Transient reflectivity measurements were made on a \Nb single crystal sample (supplied by HQ Graphene), at temperatures ranging between 15 K and 300 K. The sample was placed in a flow cryostat and cooled by liquid helium; its temperature was measured from a diode sensor in thermal contact with the sample holder. Ultrashort ($\sim$50 fs) light pulses were provided by a regenerative Ti:Sapphire amplifier, producing pulses at 800 nm (1.55 eV) and 250 kHz. TR-OS measurements with broadband detection made use of a white-light supercontinuum as a probe, extending from 0.8 eV to 2.6 eV. It is generated by focusing $\sim$1 $\mu$J/pulse energy in a 3 mm thick sapphire crystal. The detection is performed with InGaAs linear-photodiode-array (PDA) sensors with sensitivity extended in the visible spectral range. Further details on the setup used are reported in \cite{Perlangeli_2020}. Single colour probe measurements were made using light pulses produced by an infrared Optical Parametic Amplifier (OPA), tuned to 1200 nm (1.03 eV). Single-channel detection with an InGaAs pin photodiode and lock-in acquisition with a fast chopper modulation (60 kHz) ensure a signal-to-noise ratio orders of magnitude larger than broadband detection. This fact allowed us to perform scans as a function of temperature at very low fluence. The two probes share the same optical path and use the same focusing elements, hence have similar beam parameters. The pump pulse was centred at 800 nm (1.55 eV), with a duration of $\sim$50 fs, as directly provided by the Ti:Sapphire source. Measurements were made using a pump fluence of 30-45 $\mu$J/cm$^2$, unless otherwise stated, and the measured signal is denoted as $\D R/R$ and is obtained as: $\D R/R$=$(R_{pumped}-R_{unpumped})/R_{unpumped}$.

\textbf{Details of the numerical computation}

Ab-initio calculations were performed within density functional theory and the generalized gradient approximation \cite{pbe} making use of the Quantum ESPRESSO \cite{qe} code. We used ultrasoft \cite{ultrasoft} and norm conserving \cite{norm} pseudopotential for Nb and Se, respectively, a 35 Ry energy cutoff, and a 24$\times$24$\times$8 mesh of $k$ points for the electronic integrations, with an electronic smearing parameter of 0.01 Ry. The experimental lattice constants \cite{exp_lattice} $a=3.44~$\AA~ and $c=12.55~$\AA~ were adopted, while relaxing the Se atomic position along the $c$ axis until the forces were smaller than 0.005 $eV/$\AA. The joint density of states (JDOS) and the interband contribution to the optical conductivity have been calculated using a Wannier interpolation scheme \cite{wann_interp1, wann_interp2} as implemented in the Wannier90 code \cite{w90}. The dipole matrix elements are evaluated from the velocity operator $\hat{v}_\alpha=-i/\hbar [\hat{r}_\alpha, H ]$ as $d^{nm}_{\alpha\bm k} = \langle \psi_{n\bm k}\vert \hat{v}_\alpha \vert \psi_{m\bm k}\rangle$, with $\vert \psi_{n\bm k}\rangle$ being the Kohn-Sham eigenstate with band index $n$ and crystal momentum $\bm k$. A denser 60$\times$60$\times$20 mesh of $k$ points with adaptive broadening \cite{wann_interp1, wann_interp2} has been used for the calculation of JDOS and optical conductivity.

\vspace{0.5cm}
{\em ACKNOWLEDGEMENTS} 

\textbf{Funding:} The work was supported by the the Italia-India collaborative project MAECI SUPERTOP-PGRO4879, by the Sapienza University via Ateneo 2019 RM11916B56802AFE, and by Regione Lazio (L.R. 13/08) under project SIMAP. P.B. and L.B. acknowledge Italian MIUR under the PRIN project ``Tuning and understanding Quantum phases in 2D materials - Quantum2D'',  grant n. 2017Z8TS5B, and the CINECA awards under the ISCRA initiative (Grants HP10B3EDF2, HP10BSZ6LY), for the availability of high performance computing resources and support. \textbf{Author Contribution:} F.C. and F.P. conceived the experiment. F.C. coordinated the research activities with input from all the coauthors. F.C. developed the time-resolved spectroscopies. Measurements were performed by F.C. and D.T.P. The analysis of the time-resolved data was performed by F.C. and D.T.P. Calculations were performed by P.B. and L.B. The text was written by D.T.P. with major input from F.C., F.P. and L.B. All authors discussed the results and the interpretation and revised the manuscript. \textbf{Competing interests:} The authors declare that they have no competing interests. \textbf{Data and materials availability:} All data needed to evaluate the conclusions in the paper are reported in the paper and/or the Supplementary Materials. Additional data related to this paper may be requested from the authors.

\bibliography{Literature_NbSe2}

\end{document}